\documentclass{iopart}
\usepackage{graphicx}

\begin{document}
\jl{1}	

\title[First order rigidity transition in internally stressed networks]{First order rigidity transition and multiple stability regimes for random networks with internal stresses}

\author{D. A. Head}
\address{Division of Physics and Astronomy, Vrije Universiteit,
De Boelelaan 1081, Amsterdam, Netherlands}

\date{\today}

\begin{abstract}
By applying effective medium--style calculations to random spring networks, we demonstrate that internal stresses fundamentally alter the nature of the rigidity transition in disordered materials, changing it from continuous to first--order and increasing the mean coordination number $z$ at which rigidity first occurs. Furthermore, we predict the existence of a novel stability regime at low $z$ when the distribution of stresses is asymmetric. Means of verifying these predictions are suggested.
\end{abstract}



\section{Introduction}
\label{s:intro}

Predicting under what conditions a given material will support an applied load without undergoing plastic deformation is clearly of great importance to materials science and industry alike. When the material in question is ordered, the periodicity of the microstructure allows the elastic moduli to be derived from the properties of the constituent particles in a manner that is, at least in principle, straightforward. Inhomogenous materials pose significant new problems~\cite{SahimiBook}. Model disordered systems have revealed that the elastic moduli typically vanish continuously as the volume fraction (or some related parameter) drops below a critical value~\cite{CFModuli}. This {\em rigidity transition} appears to bear some of the hallmarks of continuous phase transitions in thermal systems~\cite{Jacobs95}, such as a diverging fluctuation correlation length and some degree of universality in scaling behaviour near the transition~\cite{Head03}. Experiments on materials such as chalcogenide glasses are broadly consistent with this picture~\cite{SahimiBook,Chalcogenides,Tanaka87} (but see later).

In all of the model systems considered thus far, it has been assumed that the material is initially either everywhere unstressed, or everywhere under tension~\cite{Tang87}. An overlooked possibility is the existence of {\em internal} stresses, {\em i.e.} stresses that exist on the microscopic scale that average to zero on macroscopic lengths. This omission would be valid if the material was formed in a way that allowed the total elastic energy stored in interparticle bonds to fully relax to its global minimum. However, there has been much recent interest in materials for which this is unlikely to be true. Glasses~\cite{Angell95}, particulate constructs~\cite{BouchaudProc}, and soft condensed matter systems such as dense foams and emulsions~\cite{Sollich97} are formed by the kinetic arrest of the constituent particles under suitable driving and boundary conditions, producing configurations which have a quenched--in distribution of interparticle forces. Indeed, simulations of ground--state atomic glasses have revealed internal stresses far greater in magnitude than the macroscopic value~\cite{Fynewever00}, and the long--time relaxation of a range of `jammed' soft materials has been generically attributed to stressed local regions~\cite{Cipelletti03}.

That internal stresses will alter the mechanical response is clear. Elastic filaments provide a familiar and canonical example: when placed under tension they may be plucked and will return to their original state, in contrast to compressed bands which readily buckle~\cite{LandauLifshitz}. Such considerations becomes no less important on microscopic length scales, where internal forces can significantly alter the mechanical stability of the ensemble, even in the absence of a macroscopic prestress~\cite{Alexander98}. States of self--stress are already known to alter the classic Maxwell counting rules, increasing the number of displacement modes that do not violate the constraints imposed by interactions~\cite{Calladine78}. However, it is not yet known how internal stresses alter the nature of the transition.

The purpose of this paper is to determine the role of internal stresses on the rigidity transition to and from mechanically stable states for a simple class of disordered solid, namely random Hookean spring networks at zero temperature. This is achieved using a novel, analytically--tractable approximation scheme that qualitatively agrees with known results for unstressed systems. We find that {\em any} internal stress changes the transition from continuous to first order, {\em i.e.} the elastic moduli are {\em finite} at the onset of rigidity. Furthermore, the mean number of contacts per particle $z$ at which the system becomes rigid increases with respect to its unstressed value, with a power law dependency on the magnitude of internal stresses. We also predict the existence of a new stable regime at low $z$ for asymmetric stress distributions. Means to experimentally verify these predictions are suggested.

\section{Assumptions of the model}
\label{s:assumptions}

Given that the approach adopted in this article involves several novel approximations, it is perhaps sensible to first discuss their strengths and limitations with respect to more established approaches. These will be discussed in this section. For future reference, the combined approximation scheme used here shall be referred to as the {\em mean mode approximation} or MMA.

The closest scheme to MMA is known as the {\em effective medium approximation} or EMA. Here, the inhomogeneous network is mapped onto an analogous, homogeneous substrate for which the Green's response to a point force is known. For bond--diluted lattice models, the choice of analogous system is clear: a complete lattice with every bond occupied, but where every spring has an effective spring constant~$k^{\rm eff}$. The goal is then to find $k^{\rm eff}$~\cite{Feng85}. A key problem in extending this method to prestressed systems is that the Green's function is usually not known, even for a homogeneous system~\cite{Tang87}. The MMA approach, on the other hand, instead assumes that the displacement field around a perturbed bead takes a particular, averaged form, as detailed below. This can be viewed as a more drastic approximation than EMA, and indeed the reduction in the local degrees of freedom inevitably results in the rigidity transition occuring at a lower coordination number $z$ as~EMA. However, it has the advantage of not requiring an analogous system (with a known Green's function) to be identified, and can therefore be applied to unstressed and internally stressed networks alike. For unstressed networks, for which results from EMA are known, the MMA approach adopted here gives {\em qualitatively} the same behaviour as EMA, including a transition at finite coordination number $z$ and an exponent of unity for the elastic moduli as one traverses the transition. It is therefore not unreasonable to suppose that it also qualitatively predicts the correct behaviour when internal stresses are incorporated.

Much of our understanding of rigidity percolation has come from simulations, so it is natural to ask why a numerical scheme has not been adopted. The answer is essentially one of simplicity. For unstressed networks, it is straightforward to simply dilute bonds from an ordered lattice, taking care to slightly displace the lattice nodes to avoid colinearity and coplanarity of bonds. Force balance is obeyed everywhere by the simple fact that all forces are zero. This is not true in the presence of stresses: removing a bond from a stressed network will typically violate local force balance, requiring relaxation of the network to a new stable configuration for every contact broken. Thus networks constructed by a physically realistic procedure that do not relax all of their internal stresses are fundamentally a product of their history, and CPU--intensive algorithms such as molecular dynamics will be required to generate the proper starting condition. Such methods are beyond the scope of this article. The aim of this paper is partly to inspire simulation work to verify the predictions made, but also to pave the way for a full, theoretical understanding of the rigidity transition in stressed systems, for which the MMA provides a first, quite possibly qualitatively correct, first step.

Finally, we comment on the use of central force networks. The history of rigidity percolation has been somewhat confused by the study of central force networks, although the situation for unstressed networks is now clear (see {\em e.g.}~\cite{SahimiBook} for a full range of results). An obvious objection is then, why risk jeopardising this work with a potentially troublesome system? The answer is twofold. Firstly, simplicity: central force systems have fewer parameters to consider and therefore give a clearer picture of the effects of internal stresses. Secondly, however, and much more importantly, is that many of the systems mentioned in the introduction (for which internal stresses are likely to exist), such as wet foams, emulsions, frictionless granular media and Lennard--Jones systems~\cite{Angell95,BouchaudProc,Sollich97}, {\em are} central force. Studying central force systems is therefore of immediate and significant interest if one hopes to understand these materials. It also serves as a first step towards understanding other materials which are not central force, such as frictional granular media and atomic systems, which could be tackled by constructing an extending MMA scheme with transverse forces.

\section{Definition of the model}
\label{s:model}

Our model system consists of a collection of particles $\alpha$ interacting {\em via} some known finite--ranged interaction potential to produce a static body in which force balance is  everywhere obeyed. (In the language of networks, the $\alpha$ are nodes interconnected by force--transferring bonds.) The position of each particle is specified by the $d$--dimensional Cartesian vector ${\bf x}^{\alpha}$, and the displacement to a connected particle $\beta$ is ${\bf x}^{\beta}-{\bf x}^{\alpha}=r^{\alpha\beta}{\bf{\hat n}}^{\alpha\beta}$ in terms of the unit vector ${\bf{\hat n}}^{\alpha\beta}$ and the centre--to-centre distance $r^{\alpha\beta}>0$. The force on $\beta$ due to $\alpha$ is given by the central law $f(r)$ as ${\bf f}^{\alpha\beta}=f(r^{\alpha\beta}){\bf{\hat n}}^{\alpha\beta}$, so compressive contacts correspond to positive $f$. For simplicity, we assume all interactions are Hookean with identical spring constants $k>0$ and natural lengths $r_{0}$, {\em i.e.} $f(r)=-k(r-r_{0})=-kr_{0}\varepsilon$ in terms of the dimensionless extension $\varepsilon=(r-r_{0})/r_{0}$.

The macroscopic stress in the initial configuration depends on the joint probability distribution of ${\bf{\hat n}}^{\alpha\beta}$ and $r^{\alpha\beta}$. For simplicity, we assume here that the bond vectors ${\bf{\hat n}}^{\alpha\beta}$ are uniformly distributed on the unit $(d-1)$--dimensional hypersphere, and independent of the $r^{\alpha\beta}$. This ensures the macroscopic stress field is isotropic. Since our ultimate goal is to elucidate the role of internal stresses, we look for the simplest distribution $P(\varepsilon^{\alpha\beta})$ that gives zero global stress and thus choose $P(\varepsilon)=p\delta(\varepsilon-\varepsilon_{0})+(1-p)\delta(\varepsilon-\varepsilon_{1})$ with $p\in(0,1)$. The macroscopic stress tensor is then diagonal with pressure $\sim kr_{0}^{2-d}[p\varepsilon_{0}+(1-p)\varepsilon_{1}]$ for $|\varepsilon_{i}|\ll1$, and requiring that this vanishes gives $\varepsilon_{1}\approx-\varepsilon_{0}p/(1-p)$ in terms of the two free parameters $\varepsilon_{0}$ and $p$. Note that here and below we consider $|\varepsilon_{i}|\ll1$, {\em i.e.} a small perturbation around the unstressed limit $\varepsilon_{i}\equiv0$, as this facilitates some of the subsequent analysis, although no crucial modification for larger deformations are expected. Physically, this corresponds to systems of slightly deformed particles, or to a network that has relaxed close to, but not quite reaching, its global energy minimum in which all nodes are separated by their natural spring lengths.

The mechanical stability of a configuration is determined by applying an infinitesimal external force $\delta{\bf f}^{\rm ext}$ to a randomly selected particle~$\alpha$, and allowing all particles $\beta$ (including $\alpha$) to relax into a nearby configuration ${\bf x}^{\beta}+\delta{\bf x}^{\beta}$. If this nearby configuration obeys force balance, and the work done by the external agent $\delta W$ is positive, the system is deemed stable; negative work is assumed to signify mechanical instability due to the inability of the interparticle forces to support the load. Note that this force probe need not come from some spontaneous fluctuation in the steady state, whose origin would be obscure for the athermal systems under consideration here, but may be due to mechanical noise from an external source, or the final fluctuation in contact forces as the system is quenched into its final configuration by whatever preparation procedure is employed.

In analogy with statistical mechanics, the macroscopic response is expected to depend on the microscopic configuration only through a small set of suitably selected parameters. We henceforth ensemble average over all local degrees of freedom consistent with a given bond extension distribution $P(\varepsilon)$ and the mean coordination number (number of bonds per particle)~$z$. The requirement of force balance in the perturbed configuration is

\begin{equation}
\delta f^{\rm ext}_{i}
+k
\left\langle
\sum_{\beta}
A^{\alpha\beta}_{ij}(\delta x^{\beta}_{j}-\delta x^{\alpha}_{j})
\right\rangle_{P(\varepsilon),z}
=
0
\label{e:force_balance}
\end{equation}

\noindent{}where the sum is over all $\beta$ interacting with $\alpha$. Here, the change in interaction force on $\beta$ due to $\alpha$ has been written as $\delta f_{i}^{\alpha\beta}=-kA^{\alpha\beta}_{ij}(\delta x^{\beta}_{j}-\delta x^{\alpha}_{j})$, where for the central forces under consideration here,

\begin{equation}
A^{\alpha\beta}_{ij}
=
{\hat n}^{\alpha\beta}_{i}{\hat n}^{\alpha\beta}_{j}
+\varepsilon^{\alpha\beta}
\left(
\delta_{ij}-{\hat n}^{\alpha\beta}_{i}{\hat n}^{\alpha\beta}_{j}
\right),
\label{e:Aij}
\end{equation}

\noindent{}with $\delta_{ij}$ the Kr\"onecker delta. This projects out the component of the relative displacement parallel to the bond, which alters the magnitude of the force by an amount proportial to $-k$, and the transverse component, which (in this linearised scheme) describes the rotation of the original force $-kr_{0}\varepsilon^{\alpha\beta}$.

The displacement field $\delta{\bf x}$ depends on the entire initial configuration and is too complex to treat exactly. Instead, observe that $\langle\delta{\bf x}^{\alpha}\rangle=\lambda\delta{\bf f}^{\rm ext}$ for the isotropic systems under consideration here, where the value of the compliance $\lambda$ is determined later. A first approximation is then to replace the global dependency of $\delta{\bf x}^{\alpha}$ in (\ref{e:force_balance}) by the local form $\delta{\bf x}^{\alpha}=\lambda\delta{\bf f}^{\rm ext}$. No such average form is immediately apparent for $\delta{\bf x}^{\beta}$, but closure is possible by assuming that the change in contact force $\delta{\bf f}^{\alpha\beta}$ can be viewed as an external force on $\beta$, {\em i.e.} $\delta{\bf x}^{\beta}=\lambda\delta{\bf f}^{\alpha\beta}$ with the same $\lambda$ as before. Inserting these two approximations into (\ref{e:force_balance}) and (\ref{e:Aij}), and eliminating the $\delta{\bf x}$'s gives

\begin{eqnarray}
\delta f^{\alpha\beta}_{i}
&=&
S^{\alpha\beta}_{ij}
\delta f^{\rm ext}_{j},
\nonumber\\
S^{\alpha\beta}_{ij}
&=&
\left[
1+(\lambda k)^{-1}
\right]^{-1}
{\hat n}^{\alpha\beta}_{i}{\hat n}^{\alpha\beta}_{j}
\nonumber\\
&&+
\left[
1+(\varepsilon^{\alpha\beta}\lambda k)^{-1}
\right]^{-1}
\left(\delta_{ij}-{\hat n}^{\alpha\beta}_{i}{\hat n}^{\alpha\beta}_{j}\right)
\label{e:Sij}
\end{eqnarray}

\noindent{}The second rank tensor $S^{\alpha\beta}_{ij}$ gives the propagation of force from $\alpha$ to a connected particle $\beta$, and is expressed here in terms of projection operators parallel and perpendicular to the contact vector ${\bf{\hat n}}^{\alpha\beta}$. Note that the unphysical singularity at $\varepsilon\lambda k=-1$ is avoided by the stability equation given below.

Inserting (\ref{e:Sij}) into (\ref{e:force_balance}) and averaging over $P(\varepsilon)$ and ${\bf{\hat n}}^{\alpha\beta}$ gives the following equation for $\lambda$,

\begin{equation}
d\left(1-\frac{1}{z}\right)
=
\frac{1}{1+\lambda k}
+
(d-1)
\left\{
\frac{p}{1+\varepsilon_{0}\lambda k}
+ 
\frac{1-p}{1+\varepsilon_{1}\lambda k}
\right\}
\label{e:cubic}
\end{equation}

\noindent{}The work done by the external agent is $\delta W=\frac{1}{2}\lambda\left|\delta{\bf f}^{\rm ext}\right|^{2}$, so stability corresponds to positive, real roots of $\lambda$ in (\ref{e:cubic}). Although $\lambda$ is in principle a measurable quantity, a more convenient order parameter for quantifying the order of the transition is the bulk elastic modulus $K$, which is related to $\lambda$ via $K\sim r_{0}^{2-d}\lambda^{-1}$ with a dimensionless prefactor that will not concern us here.

\section{Results}
\label{s:results}

To test the validity of the approximations, we first consider unstressed systems $\varepsilon_{0}\equiv\varepsilon_{1}\equiv0$, for which the transition is known to be continuous. In this case, (\ref{e:cubic}) reduces to a linear equation with the single solution $\lambda k=(z/d-1)^{-1}$, admitting stable systems only when $z>d$. The corresponding modulus is $K\sim kr_{0}^{d-2}(z/d-1)$, which can be written as a power law $K\sim(z-z_{\rm c})^{f}$ to explicitly demonstrate the continuous transition at $z=z_{\rm c}=d$ with an exponent $f=1$. Effective medium theory predicts the same exponent but at the higher transition point $z^{\rm ema}_{\rm c}=2d$~\cite{Feng85}, as predicted by Maxwell counting. However, EMA requires knowledge of the Green's function for an equivalent homogeneous system. The advantage of our MMA approach is that it predicts a finite--$z$ transition without requiring a known Green's response, and thus can be applied to a broader range of materials, at least qualitatively. In particular, the generalization to internal stresses is straightforward, as we now demonstrate.

\begin{figure}
{\includegraphics[width=12cm]{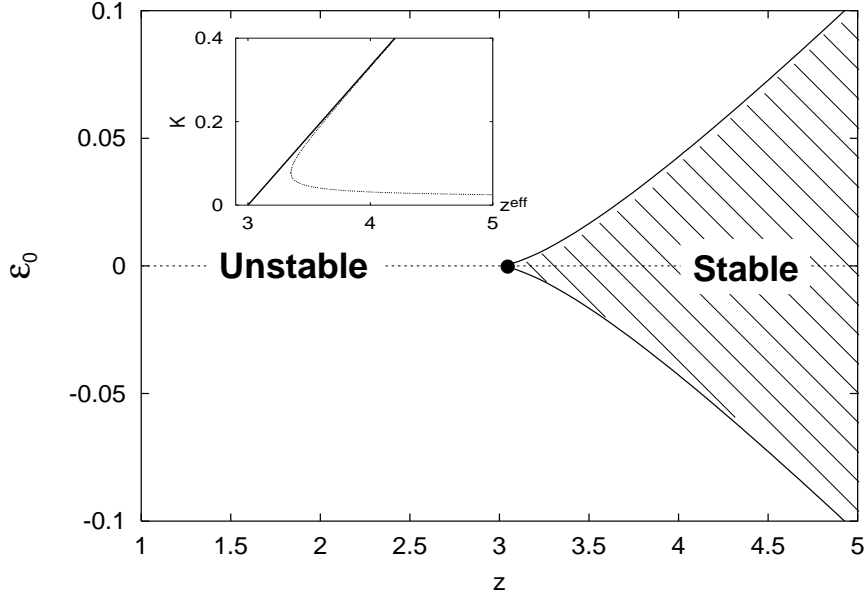}}
\caption{Combinations of coordination number $z$ and dimensionless extension $\varepsilon_{0}$ that generate stable systems, here shown for the symmetric case $p=\frac{1}{2}$ (so $\varepsilon_{1}=-\varepsilon_{0}$) and dimension $d=3$. The solid disc is the continuous transition; all other transitions are first order. {\em (Inset)} $K$ versus $z$ for $\varepsilon_{0}=0$ (solid straight line) and $\varepsilon_{0}=0.01$ (dashed), demonstrating the first--order transition and multiplicity of solutions in the latter case.\label{f:sym}}
\end{figure}

The simplest state of internal stress to consider is a symmetric distribution
$\varepsilon_{0}=-\varepsilon_{1}$, {\em i.e.} $p=\frac{1}{2}$. In this case, equation (\ref{e:cubic}) only permits real positive roots in one region of parameter space corresponding to $z\geq z_{\rm c}(\varepsilon_{0})\geq d$ as plotted in Fig.~\ref{f:sym}. The transition points lie along the curve

\begin{equation}
\varepsilon_{0}^{2}=\frac{4(z-d)^{3}}{(3d)^{3}(z-1)},
\label{e:sym_border}
\end{equation}

\noindent{}which obeys the expected symmetry under $\varepsilon_{0}\leftrightarrow-\varepsilon_{0}$. The value of $K$ at the transition is most easily expressed in terms of $z-d$,

\begin{equation}
K_{\rm  trans}\sim r_{0}^{2-d}\lambda^{-1}=\frac{2kr_{0}^{2-d}(z-d)}{3d}
\label{e:sym_modulus}
\end{equation}

\noindent{}By inspection of (\ref{e:sym_border}) and (\ref{e:sym_modulus}), it is clear that the transition is first order everywhere {\em except} at $\varepsilon_{0}=0$, {\em i.e.} internal stresses both move the transition to a higher $z_{\rm c}$, and change its nature to first order.

This result broadly holds for asymmetric distributions with $p\neq\frac{1}{2}$, {\em i.e.} $\varepsilon_{0}\neq-\varepsilon_{1}$, although the transitions for $\varepsilon_{0}>0$ and $\varepsilon_{0}<0$ are no longer symmetrical. Expanding about $z-d$ to ${\mathcal O}[(z-d)^{2}]$, the transitions are at

\begin{equation}
\varepsilon_{0}
=
\sqrt{\frac{4(1-p)(z-d)^{3}}{p(3d)^{3}(d-1)}}
+
\frac{(1-2p)(z-d)^{2}}{3pd^{2}(d-1)}
+\ldots
\label{e:asym_border}
\end{equation}

\noindent{}The leading order term $\sim(z-d)^{3/2}$ admits two roots, corresponding to the transition lines for $\varepsilon_{0}>0$ and $\varepsilon_{0}<0$. The next--to--leading order term always has the same sign (that of $1-2p$) for both roots, thus breaking the symmetry about $\varepsilon_{0}=0$. $K$ at the transition is also asymmetric, but remains always first--order for $\varepsilon_{0}\neq0$,

\begin{eqnarray}
K_{\rm trans}
&\sim&
\frac{2kr_{0}^{2-d}(z-d)}{3d}
\times
\left\{
1+\frac{1-2p}{6}
\sqrt{\frac
{3(z-d)}
{d(d-1)\,p(p-1)}}
\right\}
\label{e:asym_modulus}
\end{eqnarray}

\noindent{}plus higher order terms, where the sign of the square root is chosen to match that in (\ref{e:asym_border}).

However, the modulation of the transition curves for $z>d$ is not the only effect of asymmetric stresses; a distinct stable regime with $z<d$ also emerges, as seen in Fig.~{\ref{f:asym}. The $\varepsilon_{0}\equiv0$ boundary of this region stretches from $z=1$ to $z=z^{*}$ with

\begin{equation}
z^{*} -1 = (d-1)
\left[
1+\frac{4dp(1-p)}{(1-2p)^{2}}
\right]^{-1},
\end{equation}

\noindent{}and extends strictly in the direction of sign {\em opposite} to $1-2p$ ({\em i.e.} $\varepsilon_{0}<0$ for $p<\frac{1}{2}$), narrowing as $\varepsilon_{0}$ increases in magnitude. In words, this regime corresponds to systems with a small proportion of highly compressed contacts in a sea of weakly tensile bonds ({\em i.e.} small $p$ and $\varepsilon_{0}<0$). We speculate that the $z>d$ and $z<d$ stable regions may be analogous to the glass and gel states, respectively, in colloids with short--range attractions~\cite{Puertas02}, although the transition here is a purely percolation phenomenom and has no entropic component. That this new regime extends down to the unphysical value $z=1$ is most likely a further consequence of the approximations involved: just as the unstressed transition lies at $z=2d$ in real systems, so would the lowest allowed value be $z=2$, as realised in {\em e.g.} long strings of beads under tension.

\begin{figure}
{\includegraphics[width=12cm]{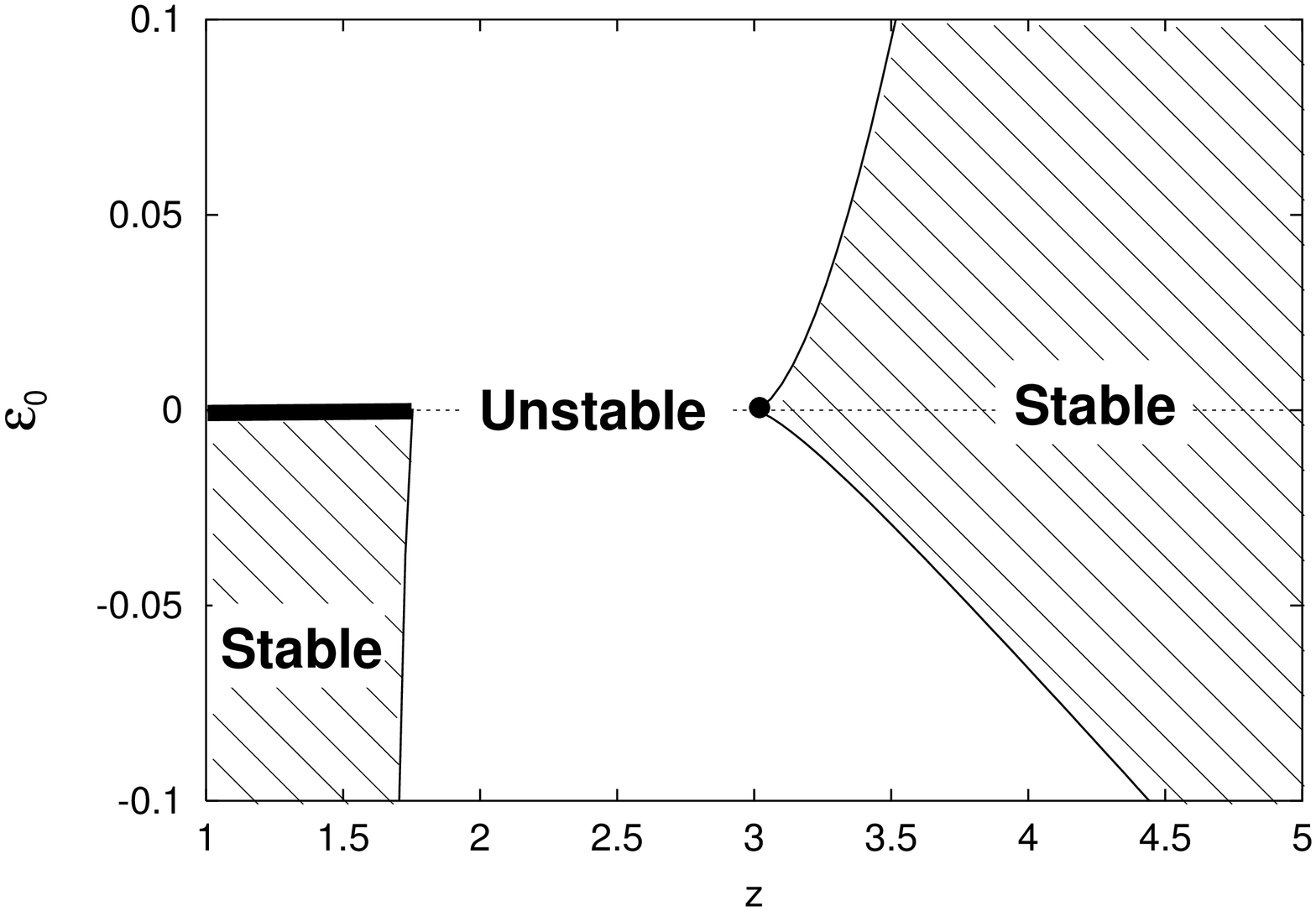}}
\caption{As Fig.~\ref{f:sym} but for an asymmetric distribution of internal stresses in which a fraction $p=0.1$ bonds have dimensionless extension $\varepsilon_{0}$ and $1-p=0.9$ have $\varepsilon_{1}=-\varepsilon_{0}p/(1-p)$. The thick line and solid disc correspond to continuous transitions; all other transitions are discrete.\label{f:asym}}
\end{figure}

\section{Discussion and summary}
\label{s:disc}

Experimentally verifying these predictions should, in principle, be straightforward. According to this theory, the magnitude of $K$ at the transition, and the increase in the transition value $z_{\rm c}$, should both increase monotonically with the magnitude of the internal stresses. Indeed, this effect may have already been observed in Si$_{x}$Se$_{1-x}$ glasses~\cite{Chalcogenides}, although it is currently interpreted as evidence for an intermediate rigid--but--stressless regime resulting from self--organization of the contact topology during cooling~\cite{Thorpe00}. Differentiating between these two descriptions could be achieved by varying the rate of quench~\cite{Tanaka87}: faster quenches will generate a broader distribution of internal stresses, and hence (according to the theory presented here) a higher $z_{\rm c}$ and greater jump in elastic properties at the transition. Slow quenches would give small internal stresses and could even appear as continuous transitions, to within experimental error. It is less clear how the degree of asymmetry $p\neq\frac{1}{2}$ will depend on sample preparation, and hence it is difficult to predict when the low--$z$ stable regime will occur. Molecular dynamics simulations, where some quantity analogous to $p$ could be directly measured, may help to resolve this issue.

Unless there exist residual interactions capable of maintaining solidity, systems placed in an unstable region will plastically rearrange according to the dynamical properties of the particles and the surrounding medium. Clearly such dynamics cannot be described by the static theory presented here. Nonetheless, some qualitative observations based on our results can be made. Firstly, unless the system {\em completely} relaxes its internal stresses during the dynamical phase, it will {\em not} have a coordination number at or near to the usually quoted percolation transition ($z=2d$ here), since such systems are simply not stable when there are internal stresses. This is a qualitative prediction that could  be verified by molecular dynamics simulations, for instance. Secondly, assuming that the rearranging system eventually comes to halt on the boundary between stable and unstable regions, as recently proposed for the clustering of weakly--attractive colloids~\cite{Kroy03}, then the elastic moduli will only become arbitrarily small {\em if the magnitude of internal stresses are similarly small}. Again, this is a clear prediction of a qualitative difference between stressed and unstressed networks.

A secondary aspect of this article has been the introduction of a new approximation method, the MMA, which predicts a range of non-trivial behaviour despite the simplicity of its assumptions. Indeed, it qualitatively reproduces the results of effective medium theory for unstressed systems with much less algebra. It is therefore sensible to ask when the approximation is expected to work, and when it might fail. The MMA proceeds by closing the force balance equations under the assumption of a parameterised form of local response. While this will always fail quantitatively, it should only {\em qualitatively} fail when the actual response is very different to the assumed form. In this instance something similar to the actual mode (if known) could be employed instead. Another potential problem is that the displacement of particles connected to the perturbed one is assumed to depend {\em only} on the change in contact force with the perturbed particle itself. Given that the mean force must decay monotonically with distance (to obey force balance), the closest bonds will be perturbed the greatest and so this seems reasonable; however, exotic modes in which the force does {\em not} decay in every direction may cause this assumption to fail. Even in such instances, it is hoped the MMA could be extended to faithfully mimic the actual response.


In summary, we have argued that internal stresses qualitatively alter the nature of the rigidity transition to configurations with non--vanishing elastic modulii, making it first order and also moving the threshold coordination number $z_{\rm c}$ to a higher value. A distinct stability regime with low $z$ was also predicted to arise when the internal stresses are asymmetrically distributed. Although only central forces have been considered here, these basic findings are expected to extend to systems with bending forces and other microscopic interactions. It is hoped that numerical and experimental verification of these claims will be forthcoming.


\section*{Acknowledgments}
The author was funded by a
European Community Marie Curie fellowship.



\Bibliography{99}

\bibitem{SahimiBook} M. Sahimi, {\em Heterogeneous Materials I: Linear Transport and Optical Properties} (Springer--Verlag, New York, 2003).

\bibitem{CFModuli} S. Feng, P. N. Sen, B. I. Halperin and C. J. Lobb, Phys. Rev. B {\bf 30}, 5386 (1984); M. Sahimi and S. Arbabi, Phys. Rev. B {\bf 40}, 4975 (1989); S. Arbabi and M. Sahimi, J. Phys. A {\bf 21}, L863 (1988).

\bibitem{Jacobs95} D. J. Jacobs and M. F. Thorpe, Phys. Rev. Lett. {\bf 75}, 4051 (1995); C. Moukarzel and P. M. Duxbury, Phys. Rev. E {\bf 59}, 2614 (1999).

\bibitem{Head03} D. A. Head, F. C. MacKintosh and A. J. Levine, Phys. Rev. E {\bf 68}, 025101(R) (2003).

\bibitem{Chalcogenides} X. Feng, W. J. Bresser and P. Boolchand, Phys. Rev. Lett. {\bf 78}, 4422 (1997); D. Selvanathan, W. J. Bresser, P. Boolchand and B. Goodman, Solid State Comm. {\bf 111}, 619 (1999).

\bibitem{Tanaka87} K. Tanaka, Phys. Rev. B {\bf 36}, 9746 (1987); K. Tanaka,
Phys. Rev. B {\bf 39}, 1270 (1989).

\bibitem{Tang87} W. Tang and M.F. Thorpe, Phys. Rev. B {\bf 36}, 3798 (1987); W. Tang and M.F. Thorpe, Phys. Rev. B {\bf 37}, 5539 (1988).

\bibitem{Angell95} C. A. Angell, Science {\bf 267}, 1924 (1995).

\bibitem{BouchaudProc} See {\em e.g.} J.--P. Bouchaud in {\em Slow relaxations and non--equilibrium dynamics in condensed matter}, eds. J.--L. Barrat {\em et al.} (Springer, Paris, 2003).

\bibitem{Sollich97} P. Sollich, F. Lequeux, P. H\'{e}braud and M. E. Cates, Phys. Rev. Lett. {\bf 78}, 2020 (1997).

\bibitem{Fynewever00} H. Fynewever, D. Perera and P. Harrowell, J. Phys.: Cond. Mat. {\bf 12}, A399 (2000); S. --P. Chen, T. Egami and V. Vitek, Phys. Rev. B {\bf 37}, 2440 (1988); T. Kustanovich, Y. Rabin and Z. Olami, Phys. Rev. B {\bf 67}, 104206 (2003).

\bibitem{Cipelletti03} L. Cipellitti, L. Ramos, S. Manley, E. Pitard, D. A. Weitz, E. E. Pashkovski and M. Johansson, Farad. Disc. {\bf 123}, 237 (2003).

\bibitem{LandauLifshitz} L. D. Landau and E. M. Lifshitz, {\em Theory of Elasticity}, 3rd ed. (Butterworth--Heinmann, Oxford, 1986).

\bibitem{Alexander98} S. Alexander, Phys. Rep. {\bf 296}, 65 (1998).

\bibitem{Calladine78} C. R. Calladine, Int. J. Sol. Struct. {\bf 14}, 161 (1978); R. Connelly and W. Whiteley, SIAM J. Disc. Math. {\bf 9}, 453 (1996); J. Mech. Phys. Solids {\bf 51}, 383 (2003).

\bibitem{Feng85} S. Feng, M. F. Thorpe and E. Garboczi, Phys. Rev. B {\bf 31}, 276 (1985).

\bibitem{Puertas02} A. M. Puertas, M. Fuchs, and M. E. Cates, Phys. Rev. Lett {\bf 88}, 098301 (2002).

\bibitem{Thorpe00} M. F. Thorpe, D. J. Jacobs, M. V. Chubynsky and J. C. Phillips, J. Non--Cryst. Sol. {\bf 266--269}, 859 (2000);  M. Micoulaut, Europhys. Lett. {\bf 58}, 830 (2002).

\bibitem{Kroy03} K. Kroy, M. E. Cates and W. C. K. Poon,
{\em cond-mat/0310566}.

\endbib
\end{document}